\documentclass[%
reprint,
superscriptaddress,
 amsmath,
 amssymb,
 aps,
]{revtex4-2}

\usepackage{graphicx}
\graphicspath{{figures/}}
\usepackage{comment}
\usepackage{dcolumn}
\usepackage{bm}
\usepackage{hyperref}
\usepackage{xfrac}
\usepackage{siunitx}
\usepackage[T1]{fontenc} 
\usepackage[dvipsnames]{xcolor}
\colorlet{ForestGreen}{Black}
\begin{document}


\title{Revealing and reducing growth-induced interfacial disorder in preferentially aligned nitrogen-vacancy centers in diamond}

\author{Cheng-I Ho}
\email[]{cheng-i.ho@pi3.uni-stuttgart.de}
 \affiliation{3rd Institute of Physics, University of Stuttgart, 70569 Stuttgart, Germany.}  
 \affiliation{IQST, Center for Integrated Quantum Science and Technology, 70569 Stuttgart, Germany.}

\author{Marina Davydova}
 \affiliation{Fraunhofer Institute for Applied Solid State Physics, Tullastr. 72, 79108 Freiburg, Germany.}  

\author{Patrik Stra\v{n}\'ak}
 \affiliation{Fraunhofer Institute for Applied Solid State Physics, Tullastr. 72, 79108 Freiburg, Germany.}  

\author{Felix Hoffmann}
 \affiliation{Fraunhofer Institute for Applied Solid State Physics, Tullastr. 72, 79108 Freiburg, Germany.}  

\author{Peter Knittel}
 \affiliation{Fraunhofer Institute for Applied Solid State Physics, Tullastr. 72, 79108 Freiburg, Germany.}  

\author{Andrej Denisenko}
\email[]{a.denisenko@pi3.uni-stuttgart.de}
 \affiliation{TTI GmbH /SQUTEC, 70569 Stuttgart, Germany}  
 
\author{J\"org Wrachtrup}
 \affiliation{3rd Institute of Physics, University of Stuttgart, 70569 Stuttgart, Germany.}
 \affiliation{IQST, Center for Integrated Quantum Science and Technology, 70569 Stuttgart, Germany.}
 \affiliation{Max Planck Institute for Solid State Research, 70569 Stuttgart, Germany.}

\begin{abstract}
Nitrogen-vacancy (NV) centers in chemical-vapor-deposition (CVD) diamond can form preferentially oriented ensembles with high sensing performance and low densities of lattice defects. Thin films of this material are a cornerstone of various imaging modalities. However, nitrogen injection needed to produce such films can transiently drive growth out of equilibrium, generating interfacial strain and spin defects that degrade NV coherence. Here, we investigate this disorder in $^{12}\text{C}$-enriched, preferentially oriented NV layers grown on (111) diamond using two nitrogen-injection procedures, combined with nanometer-scale selective plasma etching and NV spin-coherence measurements. Pulsed nitrogen injection produces a pronounced nitrogen overshoot within a 60--80 nm interfacial region, generating excessive amounts of defects.
By contrast, smooth nitrogen delivery through mass flow controllers substantially suppresses interfacial disorder, yielding coherence properties close to the theoretical limit imposed by spin-bath noise. A 50-nm NV layer is used to demonstrate proton nuclear magnetic resonance detection. This work reveals the role of interfacial disorder associated with the nitrogen-doping procedure and provides a route to growing high-quality, thin NV-doped layers for quantum-sensing applications.

\end{abstract}

\maketitle

\section*{Introduction}

\begin{figure}
    \includegraphics{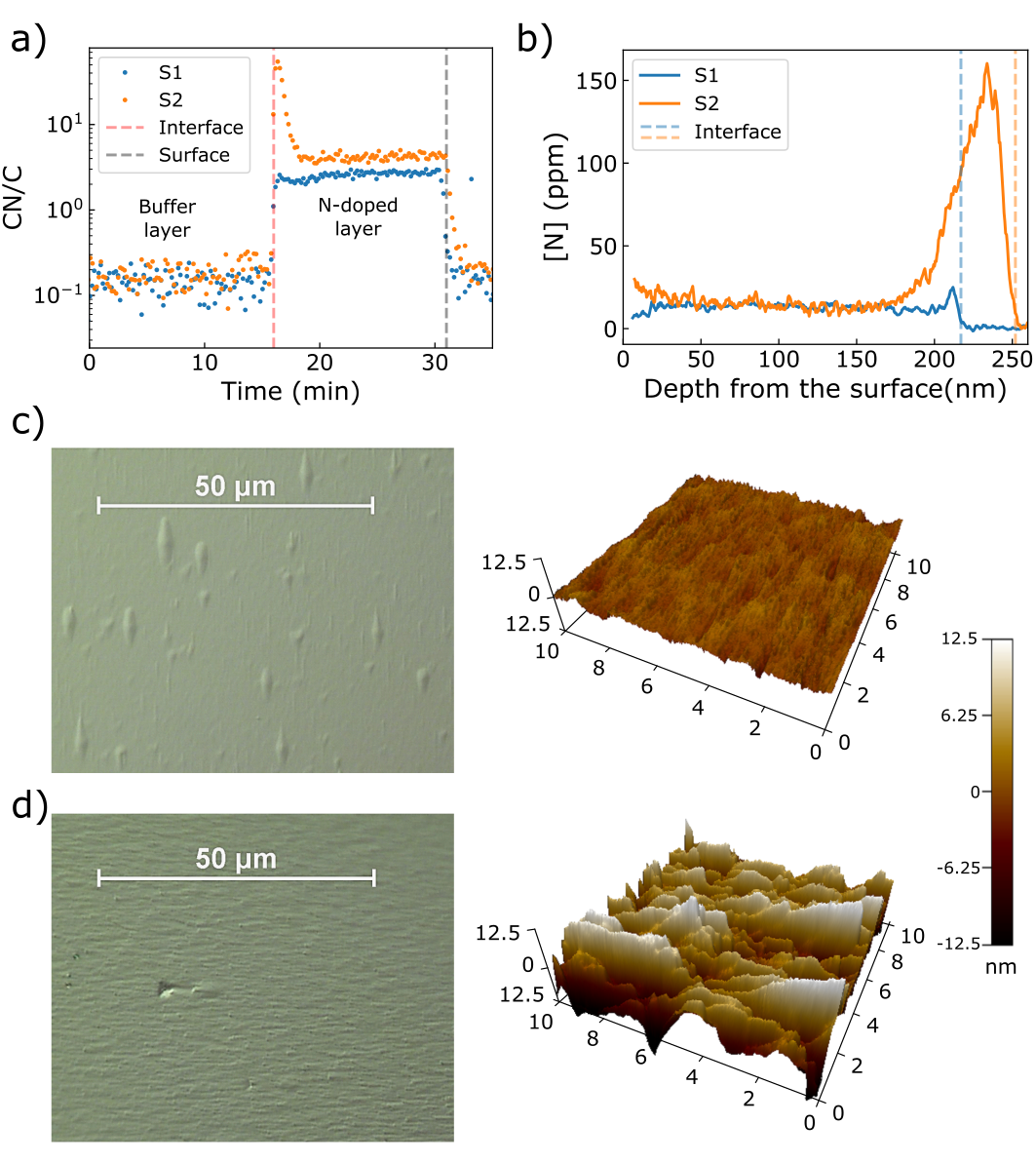}
    \caption{\label{fig1} \textbf{Nitrogen profiles and surface characterization.} 
    \textbf{a)} OES of nitrogen-containing species in the plasma during growth. The gray dashed line marks the diamond surface, and the red dashed line marks the interface between the buffer layer and the N-doped layer, corresponding to the time at which nitrogen gas is introduced.
    \textbf{b)} ToF-SIMS profiles of the nitrogen concentrations in the two samples. The blue and orange dashed lines mark the interfaces for sample S1 and S2, respectively.
    \textbf{c)} and \textbf{d)} Optical differential-interference-contrast images (left) and three-dimensional atomic force microscopy (AFM) images (right) of S1 and S2, respectively.
    }
\end{figure}

Quantum applications of solid-state defects have been extensively explored for quantum sensing, quantum computing, and quantum communication \cite{awschalomQuantumTechnologiesOptically2018}. Among the various solid-state defect platforms, the nitrogen-vacancy (NV) center in diamond has emerged as one of the most promising candidates owing to its long spin coherence times, optical initialization and readout capabilities, and coherent microwave controllability \cite{dohertyNitrogenvacancyColourCentre2013,bar-gillSolidstateElectronicSpin2013}. In particular, quantum sensing based on NV centers has attracted significant attention because it combines room-temperature operation with high sensitivity and nanoscale spatial resolution \cite{taylorHighsensitivityDiamondMagnetometer2008,bianNanoscaleElectricfieldImaging2021}.
Research has been performed for fundamental studies including the magnetic properties of two-dimensional materials \cite{jayaramProbingVortexDynamics2025, wongSupermoireSpinTextures2026} and dark matter detection \cite{chigusaNuclearSpinMetrology2025}, as well as for applications such as nano- and microscale nuclear magnetic resonance (NMR) 
\cite{loretzNanoscaleNuclearMagnetic2014a,aslamNanoscaleNuclearMagnetic2017,arunkumarMicronScaleNVNMRSpectroscopy2021,abendrothSingleNitrogenVacancyNMRAmineFunctionalized2022,sushkovMagneticResonanceDetection2014,shiSingleproteinSpinResonance2015,lovchinskyNuclearMagneticResonance2016b} and electric-field detection \cite{bianNanoscaleElectricfieldImaging2021}. Among these applications, NMR has drawn considerable attention because of its ability to detect single molecules and even single nuclear spins \cite{sushkovMagneticResonanceDetection2014,shiSingleproteinSpinResonance2015,lovchinskyNuclearMagneticResonance2016b}. A key figure of merit for a quantum sensor is its magnetic sensitivity $\eta_B$. Assuming that the target is a coherent oscillating magnetic field, the sensitivity is given by \cite{taylorHighsensitivityDiamondMagnetometer2008}:

\begin{equation} \label{eq1}
    \eta_B \propto \frac{1}{C \sqrt{n_{\text{photon}}}\sqrt{T_{\text{sens}}}},
\end{equation}
where $C$ is the contrast, $T_{\text{sens}}$ is the sensing time, and $n_{\text{photon}}$ is the number of collected photons, which is proportional to the number of quantum sensors. For decades, research has focused on improving sensitivity by optimizing these parameters.

One approach to improving all of these parameters for NV centers is chemical vapor deposition (CVD) growth. Compared with ion implantation, which produces numerous unwanted defects in the diamond lattice, CVD growth causes less lattice damage, thereby improving the coherence time and charge stability of NV centers
\cite{balasubramanianUltralongSpinCoherence2009,herbschlebUltralongCoherenceTimes2019,kimScalableNanoscalePositioning2025}. Recent work shows that the coherence time of single NV centers can be even extended to ten seconds by isotopically purifying the diamond during growth \cite{yamamotoTenSecondElectronSpinCoherence2026}.
Furthermore, by carefully tuning the growth conditions, one can achieve preferentially aligned NV ensembles \cite{michlPerfectAlignmentPreferential2014,gotzePreferentialPlacementAligned2022,osterkampEngineeringPreferentiallyalignedNitrogenvacancy2019,phamEnhancedMetrologyUsing2012}. Such an NV ensemble maintains the contrast of single NV centers while increasing the number of collected photons with numerous NV centers, thus significantly improving sensitivity. However, good coherence properties are not guaranteed. For example, when the gas composition and flow deviate from the optimal conditions, other NV orientations appear and compete with the preferred orientation \cite{jinEffectNitrogenGrowth1994,bohrInfluenceNitrogenAdditions1996,yanVeryHighGrowth2002,chayaharaEffectNitrogenAddition2004,terajiChemicalVaporDeposition2008}. The lattice disorder can also increase significantly. This produces strain and additional lattice defects, reducing the contrast and introducing additional sources of decoherence. Deviations from the optimal equilibrium conditions occur particularly when nitrogen gas is introduced into the growth chamber \cite{jinEffectNitrogenGrowth1994,bohrInfluenceNitrogenAdditions1996,yanVeryHighGrowth2002}. When a thin NV layer is grown, the NV centers are close to the interface and are therefore highly sensitive to this interfacial disorder. Although a method has been reported for obtaining a sharp interface while maintaining preferentially aligned NV centers \cite{schatzleChemicalVaporDeposition2023}, microscopic interfacial disorder on the nanoscale has not been studied in detail.

In this work, we study NV centers in diamonds grown by two different methods on (111)-oriented substrates. One sample, S1, is grown using a series of mass flow controllers (MFCs) to inject nitrogen gas smoothly, whereas the other sample, S2, is grown using pulsed injection of nitrogen gas, which causes a radical change in the growth conditions.
Spin-characterization measurements, including Ramsey, double-quantum Ramsey (DQR), spin-echo, and double electron-electron resonance (DEER) measurements, are performed to investigate sources of decoherence, including strain and spin noise. Step-etching is performed on the samples to study the properties of NV centers at different distances from the interface. Large strain is observed in sample S2, together with spin noise at the interface that is larger than the value projected from the concentration of P1 centers, indicating that the non-equilibrium growth conditions induce additional lattice defects and therefore sources of decoherence other than P1 centers.
We finally demonstrate NMR detection of protons using sample S1.
By understanding the effects of the growth conditions at the interface, this work paves the way for optimizing growth to obtain a high-quality, thin NV layer for quantum-sensing applications.

\begin{figure*}[!t]
    \centering
    \includegraphics[width=\textwidth]{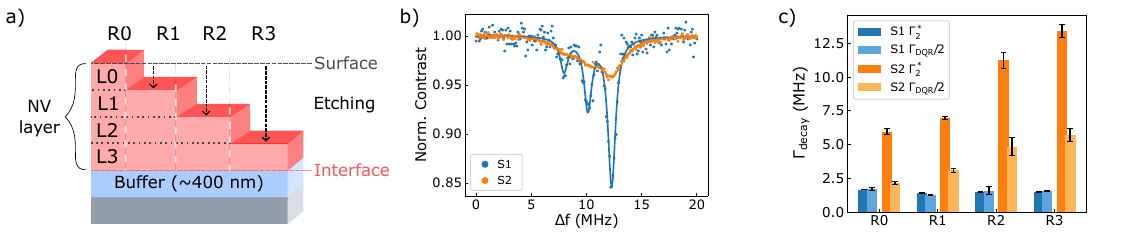}
    \caption{\label{fig2} \textbf{Step-etching and spin-coherence properties.}
    \textbf{a)} Schematic of the step-etching procedure. The measured regions are denoted R0, R1, R2, and R3 (separated by white dashed lines) and comprise different combinations of layers L0, L1, L2, and L3 (separated by gray dotted lines). The thicknesses of the NV layers in S1 and S2 are \SI{217}{nm} and \SI{252}{nm}, respectively, as determined by ToF-SIMS. The thicknesses of layers L0, L1, L2, and L3 are 110, 54, 24, and 29 nm for S1 and 110, 78, 24, and 40 nm for S2, respectively.
    \textbf{b)} Normalized ODMR spectra for S1 (blue) and S2 (orange).
    \textbf{c)} Decay rates $\Gamma_{2}^*$ and $\frac{\Gamma_{DQR}}{2}$ for S1 (blue and light blue) and S2 (orange and light orange), respectively.
    }
\end{figure*}

\section*{Results}
\noindent\textbf{CVD growth process and morphology of the samples}

Two samples, S1 and S2, are grown on identical diamond substrates with (111) surfaces in a self-built CVD system \cite{schatzleChemicalVaporDeposition2023}. Growth begins with an approximately \SI{400}{nm}-thick intrinsic diamond layer, followed by an approximately \SI{250}{nm}-thick $^{12}\text{C}$-enriched NV layer. Detailed parameters are provided in the Methods section. The only difference between the two samples is the method used to introduce nitrogen gas.
For S1, nitrogen gas is injected into the growth chamber through a series of mass flow controllers (MFCs), allowing a smooth increase in the nitrogen-gas flow and thus more stable growth conditions. For S2, nitrogen gas is injected directly, which is the commonly used but less precisely controlled method to grow N-doped diamond.
During growth, optical emission spectroscopy (OES) is used to monitor the plasma composition, as shown in Fig.\ref{fig1}(a). The intensity of the nitrogen-containing species increases slowly for S1 when nitrogen gas is introduced, whereas it increases abruptly for S2. Time-of-flight secondary-ion mass spectrometry (ToF-SIMS) is performed after growth to reveal nitrogen incorporation in the diamond, as shown in Fig.\ref{fig1}(b). Although the OES peak for S2 lasts only briefly, the sudden non-equilibrium growth conditions produce a severe overshoot of [N] in diamond that relaxes only after nearly one-third of the entire NV layer has been grown. The growth rate within this overshoot layer is estimated to be \SI{20}{nm/min}. In contrast, S1 shows only a small overshoot of [N], and its growth rate remains stable at approximately \SI{14}{nm/min}. The difference in the growth rates lead to the different overall thicknesses of the NV layers in S1 and S2, which are \SI{217}{nm} and \SI{252}{nm}, respectively.
The surface morphology of S2 (Fig.\ref{fig1}(d)) shows larger domains and a rougher surface than that of S1 (Fig.\ref{fig1}(c)), suggesting that the non-equilibrium condition caused by the nitrogen overshoot affects the growth process such that the subsequent crystalline growth retains the characteristic grain structure.

\medskip
\noindent\textbf{Analysis of decoherence sources}

When the decoherence sources affecting the NV center spins are independent, the relationship between the decay rates $\Gamma$ can be expressed as follows \cite{barrySensitivityOptimizationNVdiamond2020}:

\begin{equation} \label{eq2}
    \Gamma_{2}^* = \Gamma_{\text{spin}} + \Gamma_{\text{strain}} + \Gamma_{\text{elec}} + \Gamma_{1} + \Gamma_{\text{other}}
    \approx \frac{\Gamma_{\text{DQR}}}{2} + \Gamma_{\text{strain}}
\end{equation}

The subscripts denote the contributions from spin noise, strain, electric field, spin-lattice relaxation, and other possible dephasing sources not counted in the previous terms. It has been reported that the dephasing rate can be approximated as the sum of spin noise, strain, and electric-field noise contributions over a wide range of NV concentrations \cite{zhangUnravelingQuantumDephasing2026}. The spin-noise contribution can be determined using DQR measurements \cite{bauchUltralongDephasingTimes2018} and the factor of 2 arises from the larger phase accumulated between the NV $|-1\rangle$ and $|+1\rangle$ states during the DQR evolution time. Strain can then be assumed to be the main dephasing source other from the spin noise in a temperature-stable and magnetically homogeneous environment since $\Gamma_{\text{elec}}$ is reported only to contribute less than 10\% of the sum $\Gamma_{\text{strain}} + \Gamma_{\text{elec}}$ \cite{zhangUnravelingQuantumDephasing2026}. Furthermore, the electric-field-induced decay factor with respect to [NV] is predicted to be only \SI{11.2}{kHz/ppm^{2/3}} \cite{zhangUnravelingQuantumDephasing2026}, which is small in our samples where [NV] is always less than 1 ppm (Fig.\ref{fig4}(b)). Therefore, $\Gamma_{2}^*$ is approximated to be the sum of $\frac{\Gamma_{\text{DQR}}}{2}$ and $\Gamma_{\text{strain}}$, and the contribution from strain can be extracted.

To investigate the interfacial effects precisely, step-etching of local area is performed, enabling us to examine different distances from the interface (Fig.\ref{fig2}(a)). We first measure the optically detected magnetic resonance (ODMR) spectra of both samples, as shown in Fig.\ref{fig2}(b). Although both samples exhibit preferential NV orientation along [111] (see the Supplementary Information), their ODMR spectra differ substantially. We perform Ramsey and DQR experiments in regions R0--R3 to obtain the decay rates. The results are shown in Fig.\ref{fig2}(c).
$\Gamma_{2}^*$ is larger in every region of sample S2 than in the corresponding region of S1, which explains the broader linewidth and lower ODMR contrast of S2. Across the different regions, $\Gamma_{2}^*$ and $\frac{\Gamma_{DQR}}{2}$ in sample S1 do not show distinct variations, indicating that the spin environment throughout the entire NV layer is homogeneous. Moreover, S1 has identical values of $\Gamma_{2}^*$ and $\frac{\Gamma_{DQR}}{2}$ within the experimental uncertainties, meaning no apparent contributions of strain. In contrast, S2 shows higher decay rates, suggesting the presence of more decoherence sources than S1. It also has a larger strain contribution that increases toward the interface. This is evidence that interfacial lattice disorder creates an additional inhomogeneous environment for the NV centers and causes dephasing. Notably, this strain is still observed in measurements of R0, even after \SI{250}{nm} of overgrowth. This result is consistent with the inhomogeneous grain structure observed on the surface (Fig.\ref{fig1}(d)).

\medskip
\noindent\textbf{Spin noise analysis}

\begin{figure}
    \includegraphics{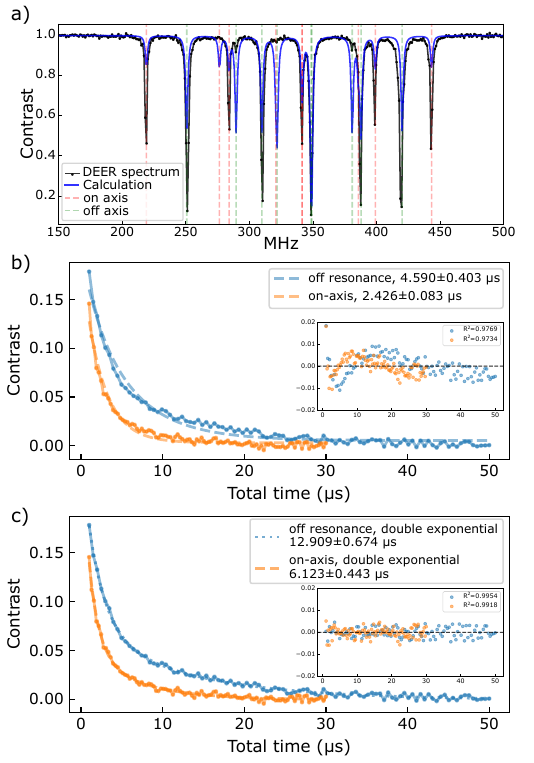}
    \caption{\label{fig3} \textbf{DEER measurements and fitting.} 
    \textbf{a)} Experimental (black) and simulated (dark blue) DEER spectra. On-axis and off-axis P1 transitions are noted with red and green dotted lines, respectively.
     \textbf{b)} Decay curves of region R0 of sample S2 when the additional $\pi$ pulse is applied at off-resonance (blue) and to the on-axis transition (orange) frequencies. Exponential functions with stretching exponent of 1--2 and 1 are used for the fits of the off-resonance and on-axis curves, respectively. The fitting residuals are shown in the inset.
     \textbf{c)} Biexponential fitting of region R0 of sample S2 and its fitting residuals.
    }
\end{figure}

It is often assumed that, in $^{12}\text{C}$-enriched diamonds with low concentration of NV centers, the main sources of spin noise are substitutional nitrogen donors, known as P1 centers \cite{hansonRoomtemperatureManipulationDecoherence2006}. Although the ToF-SIMS data provide the total nitrogen concentrations, the nitrogen atoms do not necessarily form P1 centers. Previous studies have reported that VH, NVH, NVN, vacancy chains, vacancy clusters, and interstitial-nitrogen defects can exist in diamond and also potentially produce spin noise \cite{gloverHydrogenIncorporationDiamond2004,shawImportanceQuantumTunneling2005,yamamotoExtendingSpinCoherence2013,gossInterstitialNitrogenIts2004,ashfoldNitrogenDiamond2020}. To characterize the samples in greater detail, we perform double electron-electron resonance (DEER) measurements to determine the P1-center concentrations, [P1]. Briefly, an additional $\pi$ pulse is applied during the NV spin-echo measurement to flip selected P1 centers and recouple the interaction between the NV and P1 centers, thereby causing decoherence of NV centers. Fig.\ref{fig3}(a) shows the experimental and simulated DEER spectra. 
At low to intermediate magnetic field, \SI{115}{G} in our case, some forbidden transitions are observable, leading to multiple peaks in the DEER spectrum. The simulated spectrum is obtained by calculating the transition frequencies and probabilities of P1 centers with different orientations from Jahn-Teller distortion \cite{parkDecoherenceNitrogenvacancySpin2022} and hyperfine interactions from $^{14}\text{N}$. The simulation agrees well with the experimental data, confirming that the observed peaks are indeed from P1 centers.

In DEER measurements, the additional $\pi$ pulse near \SI{218}{MHz} is chosen to resonantly flip the P1 centers oriented along the [111] axis, denoted as on-axis P1 centers, and decay rates are obtained by sweeping the interaction time $\tau$. Same experiments with the $\pi$ pulse applied at an off-resonance frequency is performed to obtain the spin-echo decay rate.
Subtracting the off-resonance decay rate from the on-axis decay rate yields the additional decay rate caused by the P1 centers, which is proportional to [P1] in a particular state and orientation \cite{bauchDecoherenceEnsemblesNitrogenvacancy2020,parkDecoherenceNitrogenvacancySpin2022}. Examples of the decay curves for sample S1 and S2 are shown in the Supplementary Information. The off-resonance curve is fitted with a stretched-exponential function. The on-axis curves, on the other hand, are fitted with simple-exponential functions since Gaussian random noise from the P1 centers leads to a Ramsey-type filter function in DEER measurements, and thus the measured curve follows an exponential decay with the exponent power of 1 when using an NV ensemble \cite{bauchDecoherenceEnsemblesNitrogenvacancy2020}. The additional decay rate caused by the P1 centers is obtained by subtracting the off-resonance decay rate from the on-resonant decay rate, from which [P1] can be calculated. Details of the fitting and the calculation of P1 concentrations are provided in the Methods section. Several measurements of different confocal points are performed in each region to obtain the statistics and determine [P1].

Although a single-exponential decay function fits all measurements in S1 well, it is not applicable to all regions of S2. The fits for R0 and R1 clearly deviate from the data, as shown in Fig.\ref{fig3}(b). For R2 and R3, however, the fit is again satisfactory (see the Supplementary Information). This is evidence that the P1-related noise sources measured in R0 and R1 of S2 are highly inhomogeneous. This is not surprising because S2 exhibits a significant nitrogen overshoot, and the [P1] near and far from the interface differ substantially, causing layer-to-layer heterogeneity. To separate the contributions from regions with different [P1], biexponential functions are used to fit the R0 and R1 data for S2, with one decay rate fixed to the result obtained for R2 of S2. Figure~\ref{fig3}(c) shows the biexponential fit to R0 of S2. The fitting residuals clearly demonstrate the fitting accuracy provided by this strategy.

\medskip
\noindent\textbf{Interfacial effects}

\begin{figure}
    \includegraphics{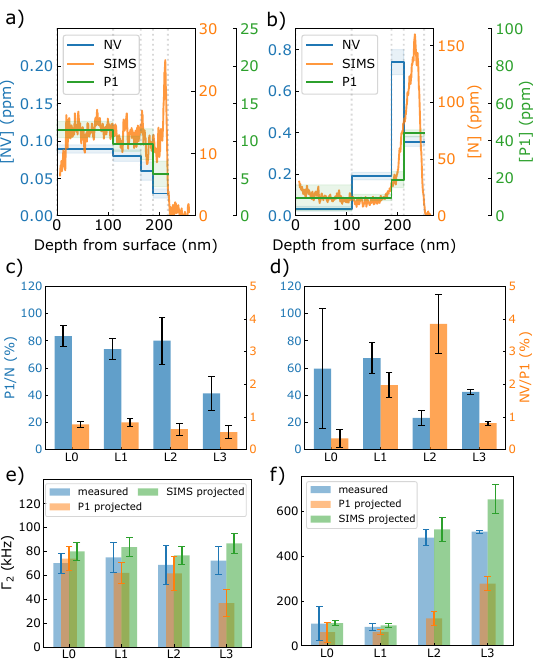}
    \caption{\label{fig4} \textbf{Summary of NV centers, P1 centers, nitrogen atoms, and spin-noise contributions in samples S1 and S2.}
    \textbf{a), b)} Concentration profiles of NV centers (blue), P1 centers (green), and nitrogen atoms (orange). The corresponding concentration axes use the same colors. The gray dotted lines separate the different layers.
    \textbf{c), d)} Conversion ratios from nitrogen atoms to P1 centers (blue) and from P1 centers to NV centers (orange).
    \textbf{e), f)} Measured $\Gamma_{2}$ (blue) and projected $\Gamma_{2}$ values calculated from the P1-center concentrations (orange) and nitrogen-atom concentrations (green). The projected $\Gamma_{2}$ values are calculated using the model reported in Ref.~\cite{bauchDecoherenceEnsemblesNitrogenvacancy2020}. Panels \textbf{a)}, \textbf{c)}, and \textbf{e)} correspond to S1; panels \textbf{b)}, \textbf{d)}, and \textbf{f)} correspond to S2. The y axes are shared for panels \textbf{a)} and \textbf{b)}, \textbf{c)} and \textbf{d)}, and \textbf{e)} and \textbf{f)}.
    }
\end{figure}

To obtain the spin-coherence properties and nitrogen-species concentrations in each NV layer, namely L0, L1, L2, and L3 in Fig.\ref{fig2}(a), we perform a weighted-average analysis. We independently determine the NV concentration in each layer from the fluorescence intensity and calculate the contribution of the NV centers in each layer to the measured spin properties. This analysis is used to calculate [P1] and the spin-coherence decay rates $\Gamma_{2}$ shown in Fig.\ref{fig4}. To examine the relationships among the concentrations, we calculate the P1-to-N and NV-to-P1 ratios, as shown in Fig.\ref{fig4}(c) and (d). We then compare the measured $\Gamma_{2}$ values with those projected using the model reported in Ref.~\cite{bauchDecoherenceEnsemblesNitrogenvacancy2020}, as shown in Fig.\ref{fig4}(e) and (f). Details of the weighted-average analysis are provided in the Methods section.

We first focus on sample S1. In Fig.\ref{fig4}(a), the [NV] and [P1] follow the same trend as [N] from layers L0 to L2. At L3, however, [NV] and [P1] decrease sharply even though [N] increases slightly because of the mild disturbance that occurs when growth of the N-doped layer begins. The P1-to-N ratios are approximately 0.8 from L0 to L2 but decrease significantly at L3, which extends approximately \SI{29}{nm} from the interface, as shown in Fig.\ref{fig4}(c). In contrast, the NV-to-P1 ratios remain stable at approximately 0.007 in all layers. This shows that the low [NV] and [P1] values in L3 result entirely from the low P1-to-N ratio, indicating inefficient conversion of nitrogen atoms into P1 centers.
Interestingly, as shown in Fig.\ref{fig4}(e), the measured $\Gamma_{2}$ for L3 coincides with the value projected from [N] rather than [P1]. 
This is evidence for the existence of other defects formed or induced by nitrogen atoms near the interface, and these defects still contribute to spin noise. One possible reason for the formation of these defects is that the sudden injection of nitrogen gas perturbs the overall gas flow and causes the growth conditions to deviate from the optimum. Because the optimal window for obtaining preferential orientation NV centers along [111] is relatively narrow, even a small disturbance in the nitrogen-gas flow can strongly affect the growth. This disturbance can promote the growth of other crystalline facets and secondary nucleation, thereby producing greater lattice disorder. 
Fortunately, this issue is not observed after the overshoot in L3, which extends over approximately \SI{29}{nm}.

Sample S2 exhibits more complicated behavior, as shown in Fig.\ref{fig4}(b) and (d). [NV] and [P1] again follow the same trend as [N] in L0, L1, and L2, whereas in L3 only [P1] increases with [N] and [NV] decreases significantly. The P1-to-N ratios are approximately 0.8 in L0 and L1, but only around 0.5 in L2 and L3, which are close to the interface. This indicates the same interfacial effect as in S1: not all nitrogen atoms are converted into P1 centers. The NV-to-P1 ratio reaches a maximum of approximately 0.04 in L2 and ranges from 0.005 to 0.01 in L0 and L3, as in S1.
We assume that L1 is close to the overshoot layer and therefore receives more vacancies from that layer during growth at $850^{\circ}\mathrm{C}$, resulting in the formation of more NV centers. L2 is at the edge of the overshoot layer, so it contains even more vacancies and has a larger NV-to-P1 ratio. However, proximity to the interface also means that more defects are formed from nitrogen atoms, resulting in the low P1-to-N ratio. In L3, the growth conditions deviate substantially from the optimum, leading to low P1-to-N and NV-to-P1 ratios. Notably, the measured $\Gamma_{2}$ agrees with the $\Gamma_{2}$ projected from [N] in L2, as shown in Fig.\ref{fig4}(f). This again suggests that nitrogen atoms that do not form P1 centers produce other sources of spin noise. The measured $\Gamma_{2}$ in L3 does not match the $\Gamma_{2}$ projected from [N]. This discrepancy may be explained by other possible spinless defects induced or formed by the large excess of nitrogen atoms, which do not contribute to spin noise \cite{ashfoldNitrogenDiamond2020}.

One notable point is that, although the interface has a profound effect on the conversion of nitrogen atoms into P1 centers, the conversion rates remain below 100\% in both samples even in layers far from the interface. This means that the other types of nitrogen-related defects are generally present in the NV layer. Various nitrogen-containing defects have been predicted and observed in the literature \cite{ashfoldNitrogenDiamond2020}.
Further investigation is needed to understand the formation of other possible defects during CVD growth.

\medskip
\noindent\textbf{Demonstration of nanoscale quantum sensing}

\begin{figure}
    \includegraphics{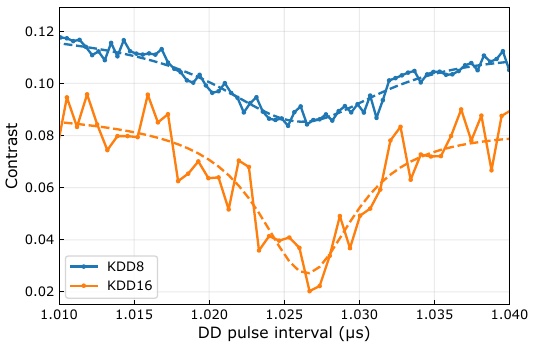}
    \caption{\label{fig5} \textbf{$^1$H NMR signal detection with S1.}
    NMR spectra measured using KDD sequences in R2 of S1. KDD8 (blue) and KDD16 (orange) denote 8 and 16 repetitions of the KDD sequence, respectively. The solid lines represent the experimental data, and the dotted lines are Lorentzian fits. The sensitivity is calculated from the two spectra.
    }
\end{figure}

We finally use region R2 of sample S1 to demonstrate quantum sensing with the thin NV layer. The robust dynamical-decoupling (DD) sequence KDD \cite{souzaRobustDynamicalDecoupling2011} is applied to detect the $^1$H NMR signal from immersion oil on the diamond surface, as shown in Fig.\ref{fig5}. The local sensitivity of the measured NV ensemble $\eta_{B_{\mathrm{rms}}}$ is \SI[parse-numbers=false]{17.66 \pm 1.35}{nT/\sqrt{\text{Hz}}}.
Using the model for detecting an oscillating magnetic field in the linear regime \cite{taylorHighsensitivityDiamondMagnetometer2008}, we also estimate the sensitivity $\eta_{B_{\text{AC}}}$ to be \SI{38.49}{nT/\sqrt{\text{Hz}}}. Both values are better than those reported for shallow implanted NV ensembles, which are typically approximately 100 nT/$\sqrt{\text{Hz}}$ \cite{tetienneSpinPropertiesDense2018a}.

The sensitivity achieved here is comparable to that of single NV centers created by CVD growth in pillars \cite{kimScalableNanoscalePositioning2025}, while using an unstructured NV-diamond and a conventional DD sequence. Moreover, when we consider the sensing ability normalized by the density of NV centers, the volume-normalized sensitivities are derived as $\eta^V_{B_{\mathrm{rms}}} = 10.81 \times 10^{-7}\text{nT}\, \text{cm}^{3/2}\, \text{Hz}^{-1}$ and $\eta^V_{B_{\text{AC}}} = 23.56 \times 10^{-7}\text{nT}\, \text{cm}^{3/2}\, \text{Hz}^{-1}$\cite{taylorHighsensitivityDiamondMagnetometer2008}. These values are comparable to reported sensitivities obtained using preferentially aligned NV ensembles in (111) diamond \cite{osterkampEngineeringPreferentiallyalignedNitrogenvacancy2019}. An additional NMR measurement using the entire 217-nm NV layer is also presented (see the Supplementary Information), showing that the entire NV layer can be used for quantum sensing.




\section*{Discussion}

We reveal interfacial disorder induced by nitrogen incorporation during CVD growth and demonstrate a mitigation strategy. The interfacial disorder induced by pulsed injection of nitrogen gas during CVD growth, which is the conventional method for growing NV diamond, contains a large amount of strain and spin noise, which are detrimental to the coherence properties of NV centers. This disorder is also found to reduce the conversion efficiency of nitrogen atoms into P1 centers, whereas the remaining nitrogen atoms that do not form P1 centers still contribute to spin noise. The resulting poor coherence properties of NV centers near the interface limit the use of such thin NV-containing layers for quantum sensing. In contrast, by carefully controlling the MFCs, a high-quality, thin NV layer with good coherence properties is obtained and enables sensitive quantum sensing. A sensitivity of \SI{17.66}{nT/\sqrt{\text{Hz}}} is achieved using a conventional DD sequence in a NMR measurement.

Although NV-NMR sensitivities as low as $\text{pT}/\sqrt{\text{Hz}}$ have been reported \cite{barryOpticalMagneticDetection2016,arunkumarMicronScaleNVNMRSpectroscopy2021}, such sensitivities are usually achieved using large NV ensembles at the microscale.
Those measurements do not have the ability to resolve nanoscale information, and often require extra microwave engineering to achieve a homogeneous $B_1$ field. Furthermore, the volume-normalized sensitivity of this work is better than those using microscale NV ensembles \cite{barryOpticalMagneticDetection2016,arunkumarMicronScaleNVNMRSpectroscopy2021} due to the confinement of the NV layer to the nanoscale.

To further improve the sensitivity, increasing the photon collection efficiency or the number of NV centers is essential. 
A recent work has demonstrated that a pillar structure with a preferentially aligned NV ensemble enables the spin-projection-noise limit by greatly increasing the collected photons, thereby further improving sensitivity \cite{maierReadoutSolidState2026}. The number of NV centers can be increased by improving the relatively low P1-to-NV conversion rate. Ultraviolet irradiation during growth \cite{fuhrmannPhotoactivationNVCenters2026} or post-growth annealing \cite{osterkampEngineeringPreferentiallyalignedNitrogenvacancy2019} may increase the NV yield, although at the cost of losing preferential alignment. Another approach is to use advanced sensing schemes for NV ensemble, such as the rectification protocol \cite{maierEfficientDetectionStatistical2025b}, which can improve the signal-to-noise ratio and reduce the measurement time. 

\section*{Methods}
\medskip
\noindent\textbf{NV diamond preparation and surface characterization}


Diamond films were synthesized in a home-built microwave plasma-assisted chemical vapor deposition (MPCVD) reactor operating at \SI{2.45}{GHz}, specifically optimized for the homoepitaxial growth of high-purity diamond. The system is equipped with a load-lock chamber featuring horizontal sample transfer and a vertical substrate lift mechanism, which together enable the formation of sharp interfaces between successive epitaxial layers.

NV-containing diamond films of nanometer-scale thickness were deposited on (111)-oriented high-pressure high-temperature (HPHT) type IIa diamond substrates. Prior to growth, the substrates were treated by inductively coupled plasma reactive ion etching (ICP-RIE) in a hydrogen-oxygen plasma in order to remove polishing-induced subsurface damage.

The samples, S1 and S2, were fabricated in a two-step deposition process. In the first step, an intrinsic buffer layer was grown using methane with natural isotopic abundance ($\text{CH}_4$) as the carbon precursor. In the second step, a nitrogen-doped layer was deposited on top using isotopically enriched $^{12}\text{CH}_4$ (99.99\%) to suppress decoherence arising from $^{13}\text{C}$ nuclear spins. Both layers were grown at a carbon gas-phase concentration of 0.5\%, whereas the nitrogen-doped layer was additionally supplied with nitrogen at an N-to-C gas-phase ratio of 40,000 ppm. During all nitrogen-containing deposition runs, the substrate temperature was maintained at $850^{\circ}\mathrm{C}$ and continuously monitored using a two-color pyrometer (R-CZQW, Chino, Japan). OES and ToF-SIMS measurements were performed during and after growth, respectively, for both samples. Further details are provided in the Supplementary Information.

Step-etching was performed by creating masks on the diamond surfaces and subsequently etching the exposed areas with a soft oxygen plasma in a reactive-ion etcher. After etching, the diamonds were boiled in a 1:1:1 (v/v) solution of sulfuric acid, nitric acid, and perchloric acid to remove surface contamination. The etching depth was measured using a DektakXT stylus profilometer (Bruker). Surface morphology was imaged using an Olympus optical microscope, and AFM profiles were acquired in contact mode over a 10 $\mu$m $\times$ 10 $\mu$m area using an Asylum Research MFP-3D instrument (Oxford Instruments).

\medskip
\noindent\textbf{Experimental setup}

A \SI{514}{nm} laser is focused through optical components and an optical fiber, switched by an acousto-optic modulator (AOM), and reflected by a dichroic mirror toward an objective. An oil-immersion objective is used to excite the sample and collect the fluorescence. The emitted NV-center fluorescence passes through the objective, a \SI{640}{nm} long-pass filter, and a 50 $\mu$m pinhole for confocal detection. The photons are detected by two avalanche photodiodes (APDs). A piezoelectric stage with a travel range of 200 $\mu$m $\times$ 200 $\mu$m $\times$ 20 $\mu$m is used to scan the fluorescence map. The microwave signal used to control the NV centers and the radio-frequency signal used to control the P1 centers are generated by an arbitrary waveform generator, amplified by separate amplifiers, combined in a diplexer, and then guided to a 50 $\mu$m copper wire lying across the diamond surface. A cubic magnet is positioned above the diamond sample to generate a $B_0$ field of \SI{115}{G} along the NV-center axis.

\medskip
\noindent\textbf{DEER fitting and [P1] calculation}

The single-exponential fit uses the function $f(\tau) = A \exp(-(\tau / T_d)^p) + C$. The biexponential fit uses the function $f(\tau) = A_1 \exp(-(\tau/T_{d1})^{p}) + A_2 \exp(-(\tau/T_{d2})^{p}) + C$. The exponent $p$ is a free parameter for the spin-echo fits and is fixed at 1 for the DEER fits. Simulated annealing \cite{tsallisGeneralizedSimulatedAnnealing1996} is used to find the global minimum, and results with fitting errors greater than 30\% are discarded. In the biexponential fits to R0 and R1 of S2, $T_{d2}$ is fixed to the values obtained from single-exponential fits to R2 of S2: \SI[parse-numbers=false]{2.130 \pm 0.035}{\mu s} for spin echo and \SI[parse-numbers=false]{1.186 \pm 0.014}{\mu s} for DEER (see the Supplementary Information). The fitted value of $T_{d1}$ is obtained for R0* and R1*, which represent L0 + L1 and L1, respectively.

The additional decay rate $\Gamma_{P1}$ caused by P1 centers is obtained by subtracting the spin-echo decay rate from the DEER decay rate. The DEER measurements are performed by applying an additional $\pi$ pulse near \SI{218}{MHz} during the electron spin echo. This frequency is resonant with one transition of the on-axis P1 centers, so the measured decay rate $\Gamma_{\text{DEER, on-axis}}$ represents only 1/12 of the total P1 spin bath. The spin-echo decay rate, denoted by $\Gamma_{\text{DEER, off-resonance}}$, is measured with an additional $\pi$ pulse applied off-resonantly at \SI{150}{MHz} to mitigate the possible interference of the additional RF pulse in the DEER experiments.
The effective decay rate from the whole P1 spin bath at \SI{115}{G} is \SI{132}{kHz/ppm} \cite{zhangUnravelingQuantumDephasing2026}; therefore, [P1] is calculated as $[P1] = (\Gamma_{\text{DEER, on-axis}} - \Gamma_{\text{DEER, off-resonance}}) \cdot 12 / 132$ ppm. The uncertainty in [P1] is propagated from the fitting uncertainty in $\Gamma_{\text{DEER, on-axis}}$ and $\Gamma_{\text{DEER, off-resonance}}$.

\medskip
\noindent\textbf{Weighted average analysis}

Assuming the noise sources in each layer are independent and the decay rate $\Gamma_{T2_i}$ of the NV centers in each layer $L_i$ is similar, the overall decay rate $\Gamma_{2}$ in a region $Rk$ can be approximated as the weighted average of the decay rates in each layer. The same assumption is applied to [P1]. The weighted average analysis is performed according to the equation
\begin{equation} \label{eq3}
\text{Data}_{Rk} = \frac{\sum_{i=k}^{j} \text{Data}_{\text{L}i} \cdot w_i}{\sum_{i=k}^{j} w_i},
\end{equation}
where $w_{i}$ is the weight of the contribution from layer $i$, determined by the relative number of NV centers in that layer. The relative number of NV centers is calculated from the fluorescence, thickness, and NV$^-$/NV$^0$ ratio of each layer (see the Supplementary Information). We use $j=3$ for all S1 data and for R2 of S2, but $j=1$ for R0* and R1* of S2 because the R2 contribution has already been separated by the biexponential fit. This analysis is applied to calculate [P1] and $\Gamma_{2}$. Error propagation is included in all calculations.

\medskip
\noindent\textbf{Estimation of sensitivities and depths of NV centers}

The conventional formula to estimate the sensitivity of an NV ensemble detecting an oscillating magnetic field with a linear response is given by \cite{taylorHighsensitivityDiamondMagnetometer2008}:

\begin{equation} \label{eq_bac}
    \eta_{B_{\text{AC}}} = \frac{\pi}{2} \frac{1}{\gamma_e C_0 \exp(-(\frac{T_{\text{sens}}}{T_{\text{decay}}})^p) \sqrt{n_{\text{photon}}}} \sqrt{\frac{T_{\text{total}}}{T_{\text{sens}}}},
\end{equation}
where $\gamma_e$ is the gyromagnetic ratio of the electron, $C_0$ is the Rabi contrast, $T_{\text{sens}}$ is the sensing time, $T_{\text{decay}}$ is the decay time, $p$ is the stretching power of the exponential decay function which is related to the decay dynamics, $n_{\text{photon}}$ is the number of collected photons, and $T_{\text{total}}$ is the total measurement time. Assuming operation under optimal conditions, with $T_{\text{sens}} = T_{\text{decay}}$, and substituting all experimental parameters, the sensitivity is calculated as $\eta_{B_{\text{AC}}} = \SI{38.49}{nT/\sqrt{\text{Hz}}}$.

On the other hand, when we consider the local sensitivity of NV ensemble detecting the statistical polarization of nuclear spins, the root-mean-square magnetic-field fluctuation $B_{\text{rms}}$ generated by the target spins is considered. In this case, the sensitivity is defined as \cite{barrySensitivityOptimizationNVdiamond2020}:

\begin{equation} \label{eq4}
    \eta_{B_{\text{rms}}} = \frac{\sigma_{\text{signal}}}{ \lvert \frac{\partial \text{ Signal}}{\partial B_{\text{rms}}} \rvert} \sqrt{T_{\text{total}}},
\end{equation}
where $\sigma_{\text{signal}}$ is the standard deviation of the signal, which is $\sqrt{n_{\text{photon}}}$ in the photon-shot-noise limit. 

The volume-normalized sensitivity is calculated by multiplying the sensitivity by square root of the confocal volume, estimated as $150^2 \pi \times t_{\text{NV}}\, \text{nm}^3$, assuming a confocal-spot diameter of \SI{300}{nm}, where $t_{\text{NV}}$ is the NV-layer thickness in nanometers \cite{taylorHighsensitivityDiamondMagnetometer2008}.

We calculate $B_{\text{rms}}$ for a semi-infinite nuclear-spin bath above the diamond and NV centers located at an effective depth $d_{\text{NV}}$ below the surface \cite{abendrothSingleNitrogenVacancyNMRAmineFunctionalized2022, phamNMRTechniqueDetermining2016}:



\begin{equation} \label{eq5}
B_{\mathrm{rms}}^{2} = \frac{\gamma_I^2 \hbar^2 \mu_0^2 \rho}{256\pi d_{\text{NV}}^3}
\end{equation}

The measured normalized contrast $C(\tau)$ under this $B_{\text{rms}}^{2}$ is

\begin{equation} \label{eq6}
C(\tau) = \exp\left(\frac{-2}{\pi^2} \gamma_{e}^{2} B_{\text{rms}}^{2} K(N \tau)\right)
\end{equation}

In these equations, $\gamma_I$ is the gyromagnetic ratio of the target spins; $\rho$ is the spin density; $\tau$ is the DD pulse interval; $K(N \tau)$ is the function determined by the DD sequence; and $N$ is the number of pulses.
More detailed derivations are provided in the Supplementary Information.

\subsection*{Data availability}
The source data related to Figs.1 --5  are publicly available at DaRus \cite{DARUS-6396_2026}.

\subsection*{Acknowledgments}
We acknowledge funding from EU via project C-QuEnS and of the BMFTR through the cluster for future QSENS and the projects QMED-NVEPR, Q4KMU2 and NeuroQ as well as DiaQNOS and QRN. We thank the Carl-Zeiss-Stiftung via QPhoton Innovation Projects and the Center for Integrated Quantum Science and Technology (IQST) for their support.

\subsection*{Author contributions}
A.D. and J.W. conceived the project. C.H. and A.D. discussed the data and prepared the manuscript with input from all co-authors. C.H., M.D., P.K. and A.D. prepared the diamond samples. C.H. performed the main measurements and data analyses. P.S. performed the ToF-SIMS measurements. F.H. performed the OES measurements.

\section*{Competing interests}
The authors declare no competing interests.

\bibliographystyle{naturemag}
\bibliography{references}
\end{document}


\setcitestyle{numbers}
\maketitle

\section*{Supplementary Note 1: OES and ToF-SIMS}
\subsection*{OES}
Plasma composition was monitored by optical emission spectroscopy using an Andor Shamrock SR-750-A-R spectrometer (LOT, UK). Emission was collected through a multi-fiber optical cable mounted at an opening in the ellipsoidal reflector near the reaction chamber and coupled via a defocusing lens to maximize the field of view.

Nitrogen incorporation was monitored through the CN emission band at \SI{288.22}{nm} ($B^2\Sigma - X^2\Pi$ violet system) \cite{vandeveldeOpticalEmissionSpectroscopy1996}. The nitrogen signal was taken to be proportional to the maximum CN-band intensity, $I_{\mathrm{CN}}$. Spectral-background correction was performed using asymmetrically reweighted penalized least-squares smoothing \cite{baekBaselineCorrectionUsing2015}, with a smoothness parameter of $10^8$ and a termination criterion of $10^{-6}$. Spectra were acquired every \SI{10}{s}, with each spectrum consisting of 10 accumulations of \SI{1}{s}.

The detection system comprised an Andor Newton CCD camera (DU940P) cooled to $-70^{\circ}\mathrm{C}$ and an Andor Shamrock 750 spectrometer with a focal length of \SI{75}{cm} and a \SI{2400}{l/mm} holographically blazed grating, yielding a spectral resolution of \SI{0.02}{nm}.

\subsection*{ToF-SIMS}
The nitrogen concentrations and layer thicknesses were determined by time-of-flight secondary ion mass spectrometry (ToF-SIMS) using an M6 Plus instrument (IONTOF GmbH, Germany). Depth profiling was performed at the center of each sample under high-vacuum conditions, with the instrument operating at a base pressure of approximately $5 \times 10^{-10}$ mbar in dual-beam mode. A 2 keV $\text{Cs}^+$ sputter beam was combined with a 30 keV $\text{Bi}^+$ primary-analysis beam. The sputter beam was rastered over an area of 150 $\times$ 150 $\mu \text{m}^2$, whereas the analysis beam was restricted to a central 50 $\times$ 50 $\mu \text{m}^2$ region to reduce crater-edge effects. Data were acquired in negative-ion mode, and charge compensation was achieved using an electron flood gun. The secondary ion signals were quantified using the relative sensitivity factor (RSF) method. The nitrogen RSF was established using an ion-implanted reference sample, and the nitrogen concentration was evaluated from the $^{12}\text{C}^{14}\text{N}^{-}$ signal. Layer thicknesses were obtained from crater depth measurements performed with the integrated scanning probe microscope (SPM).

\section*{Supplementary Note 2: Preferential orientation of the samples}

\begin{figure}[ht]
	\centering
    \includegraphics{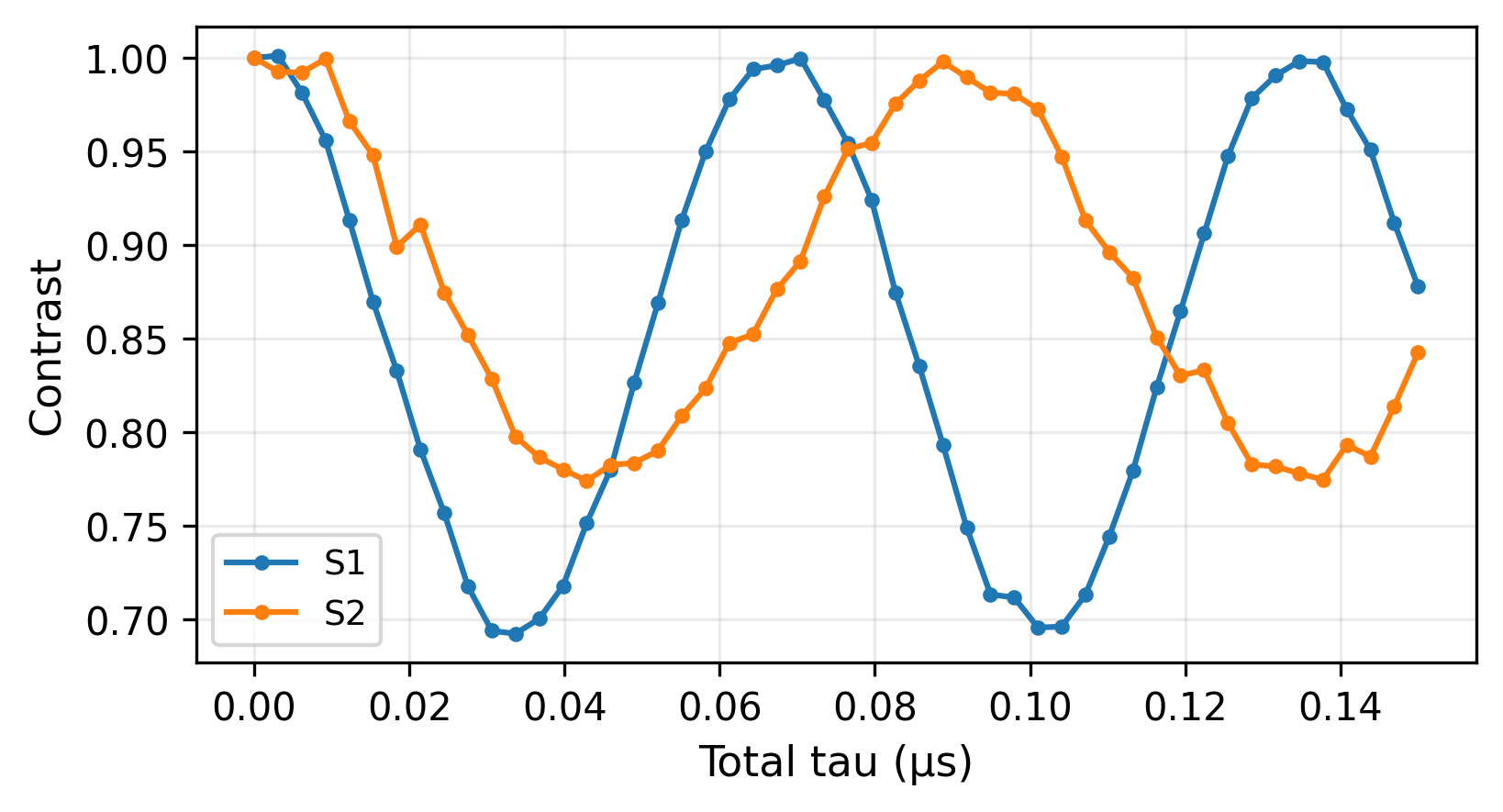}
	\caption{\textbf{Rabi oscillations.}
    Rabi oscillations of samples S1 (blue) and S2 (orange).
    }
	\label{figS1}
\end{figure}

The energy-level degeneracies in both samples, S1 and S2, are lifted by an external magnetic field $B_0$ of \SI{115}{G} applied along the NV axis. We confirm the preferential NV orientation by scanning the full spectrum, in which shows no evidence of other orientations. A strong $B_1$ field is applied in the Rabi experiments to drive all nitrogen hyperfine transitions between the electron states $|0\rangle$ and $|-1\rangle$. The observed contrast ranges from 30\% to 40\% (Fig.~S\ref{figS1}) for sample S1, consistent with the typical contrast of single NV centers, confirming that the NV centers are preferentially oriented. The lower contrast and the clearly visible decay in S2 arise from the dephasing, which is evident from the very short $T_2^*$.


\section*{Supplementary Note 3: DEER experiments}

\begin{figure}[ht]
	\centering
    \includegraphics{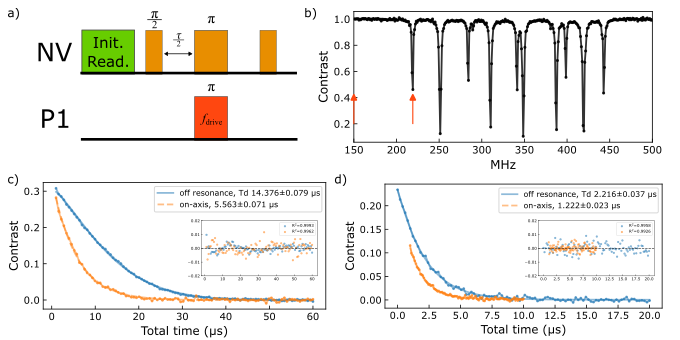}
	\caption{\textbf{DEER measurements.}
    \textbf{a)} Pulse sequence used in the DEER experiments. The additional $\pi$ pulse is applied to recouple P1-center noise to the NV centers. $f_{\text{Drive}}$ is swept to obtain the DEER spectra, and $\tau$ is swept to obtain the decay curves.
    \textbf{b)} DEER spectrum of sample S1 in R0. The red arrows indicate the values of $f_{\text{Drive}}$ shown in \textbf{a)}: \SI{150}{MHz} for the off-resonance reference and \SI{218}{MHz} for the on-axis P1 driving.
    \textbf{c), d)}
    Single-exponential fits to the spin-echo (blue) and DEER-decay (orange) curves for \textbf{c)} sample S1 in R0 and \textbf{d)} sample S2 in R2. The fitting residuals are shown in the insets.
    }
	\label{figS2}
\end{figure}

The spin-echo and DEER pulse sequences are shown in Fig.~S\ref{figS2}(a). In the DEER experiments, the frequency of the additional $\pi$ pulse is set near \SI{218}{MHz} to recouple the noise from P1 centers to the NV centers, whereas the off-resonance frequency \SI{150}{MHz} is applied as the spin-echo reference. The frequency near \SI{218}{MHz} corresponds to the transition between the states $|\pm {\frac{1}{2}}_{e},+1_{n}\rangle$ of on-axis P1 centers and agrees well with the calculation, as shown in Fig.~\ref{fig3} of the main text. The measured decay curves in the DEER measurements are fitted with either single-exponential functions, as shown in Fig.~S\ref{figS2}(c) and (d), or biexponential functions. The fitting residuals are shown in the insets. The biexponential fit is only used for R0 and R1 of sample S2, where the DEER decay is dominated by two different decay rates. The decay rates $\Gamma_{\text{DEER, on-axis}}$ and $\Gamma_{\text{DEER, off-resonance}}$ are obtained from the fits to the DEER decays with on-axis and off-resonance driving, respectively. The additional decay from the NV-P1 coupling is derived from the two decay rates, $\Gamma_{\text{P1, on-axis}} = \Gamma_{\text{DEER, on-axis}} - \Gamma_{\text{DEER, off-resonance}}$. Because there are three nitrogen hyperfine lines and four P1-center axes, including one on-axis and three off-axis orientations, the measured DEER decay $\Gamma_{\text{P1, on-axis}}$ represents only 1/12 of the coupling to the full P1 spin bath. $\Gamma_{\text{P1, on-axis}}$ is therefore multiplied by 12 and divided by the coupling coefficient \SI{132}{kHz/ppm} to obtain the total [P1]. The coefficient \SI{132}{kHz/ppm} includes the correction for forbidden transitions of P1 centers at the relatively low magnetic field, approximately \SI{115}{G} in all the experiments. A full derivation of the field-dependent coupling coefficient is provided in Ref.~\cite{zhangUnravelingQuantumDephasing2026}.

\section*{Supplementary Note 4: Parameters for weighted averaging}

The weighted-average factor $w_i$ for each NV layer is derived from the relative number of NV centers and is calculated using the fluorescence, thickness, and NV$^-$/NV$^0$ ratio of that layer. All parameters for each layer of samples S1 and S2 are listed in Supplementary Table~\ref{tableS1}. Error propagation is applied through all the calculations.

\begin{table}[htbp]
    \centering
    \caption{Calibration and layer parameters used for calculating weighted averages.}
    \label{tableS1}
    \begin{tabular}{lcccc}
        \hline
        \multicolumn{5}{l}{\textbf{Calibrated sample}} \\
        \hline
        Parameter & Value & & & \\
        \hline
        Concentration (NV / $\text{cm}^2$) & $6\times10^{12}$ & & & \\
        NV fluorescence (kcps) & 5100 & & & \\
        Laser power ($\mu$W) & 1.9 & & & \\
        Correction factor & 1.226 & & & \\
        NV$^-$/NV$^0$ & 0.479 & & & \\
        \hline
        \multicolumn{5}{l}{\textbf{Sample S1}} \\
        \hline
        Region & R0 & R1 & R2 & R3 \\
        \hline
        Thickness (nm) & 217 & 107 & 53 & 29 \\
        NV fluorescence (kcps) & 1974 & 820 & 301 & 115 \\
        Laser power ($\mu$W) & 26.4 & 25.8 & 26.8 & 27.0 \\
        Correction factor & 1.084 & 1.030 & 1.012 & 1.002 \\
        NV$^-$/NV$^0$ & 0.941 & 0.941 & 0.941 & 0.941 \\
        Normalized $w_{i}$ & 0.596 & 0.264 & 0.087 & 0.053 \\
        \hline
        \multicolumn{5}{l}{\textbf{Sample S2}} \\
        \hline
        Region & R0 & R1 & R2 & R3 \\
        \hline
        Thickness (nm) & 252 & 142 & 64 & 40 \\
        NV fluorescence (kcps) & 5282 & 5000 & 3734 & 1841 \\
        Laser power ($\mu$W) & 27.1 & 27.0 & 27.5 & 27.6 \\
        Correction factor & 1.233 & 1.222 & 1.159 & 1.077 \\
        NV$^-$/NV$^0$ & 0.967 & 0.954 & 0.942 & 0.918 \\
        Normalized $w_{i}$ & 0.071 & 0.291 & 0.354 & 0.284 \\
        \hline
    \end{tabular}
\end{table}

The measurement results for regions R0 and R1 of sample S2 are fitted with biexponential functions to separate the contribution of the overshoot region R2 from those of the remaining layers. In the biexponential fit, one decay time $T_{d2}$ is fixed to the values obtained from single-exponential fits to R2 of S2: \SI[parse-numbers=false]{2.019 \pm 0.036}{\mu s} for spin-echo and \SI[parse-numbers=false]{1.194 \pm 0.012}{\mu s} for DEER measurement.

\section*{Supplementary Note 5: Sensitivity and effective depth of the NV ensemble}

At the nanoscale, the NMR signal arises predominantly from statistical polarization of the target nuclear spins. The time-averaged magnetic field vanishes, $\langle B(t)\rangle=0$, whereas its variance and temporal correlation remain finite. During each dynamical decoupling (DD) measurement, the NV electronic spin accumulates a phase

\begin{equation}
\Delta\phi
=
\gamma_e
\int_{0}^{t_{\mathrm{sens}}}
y(t)B(t)\,dt ,
\end{equation}
where $\gamma_e$ is the electron-spin gyromagnetic ratio in $\text{rad s}^{-1} \text{T}^{-1}$ and $y(t)=\pm1$ is the modulation function generated by the DD sequence.
This sensitivity therefore does not follow the linear model used for a coherent oscillating field. Instead, the local magnetic-field sensitivity is defined as the smallest detectable magnetic field change within a measurement time:

\begin{equation}
    \label{eq:incoherent}
    \eta_B = \frac{\sigma_{\text{signal}}}{ \lvert \frac{\partial \text{ Signal}}{\partial B} \rvert} \sqrt{T_{\text{total}}},
\end{equation}
where $\sigma_{\mathrm{signal}}$ is the standard deviation of one complete measurement and $T_{\mathrm{total}}$ is its duration. In the following, we derive the sensitivity using this definition.

\subsection*{Statistical-polarization DD signal}

Two otherwise identical DD measurements are performed with opposite phases of the final $\pi/2$ pulse to obtain a two-sided differential contrast. The corresponding photon counts are denoted by $F_0$ and $F_1$, respectively. For the sensitivity calculation, the signal is defined as

\begin{equation}
    S(\tau) = F_0(\tau) - F_1(\tau)
    \label{eq:differential_signal}
\end{equation}

Assuming that $F_0$ and $F_1$ are independent Poisson-distributed photon counts, their difference follows a Skellam distribution. The photon-shot-noise standard deviation is therefore

\begin{equation}
    \sigma_{\mathrm{signal}}
    =
    \sqrt{F_0+F_1}
    \label{eq:skellam_noise}
\end{equation}




For detecting statistically polarized spin-$1/2$ nuclei with a zero-mean Gaussian distribution of accumulated phases, the normalized NV coherence is \cite{phamNMRTechniqueDetermining2016}

\begin{equation}
    C_{\mathrm{DD}}(\tau)
    =
    \exp\left[
    -\frac{1}{2}
    \left\langle
    \Delta\phi^2(\tau)
    \right\rangle
    \right]
    =
    \exp\left[
    -A(\tau)B_{\mathrm{rms}}^2
    \right]
    \label{eq:dd_coherence}
\end{equation}

with 

\begin{equation}
    A(\tau)
    =
    \frac{2}{\pi^2}
    \gamma_e^2 K(N\tau),
    \label{eq:A_definition}
\end{equation}
where $A(\tau)$ is introduced for notational convenience, $B_{\mathrm{rms}}$ is the root-mean-square magnetic field at the NV center, $N$ is the number of refocusing pulses, and $K(N\tau)$ is the DD filter functional. Assuming an infinitely narrow nuclear resonance and ideal instantaneous DD pulses, $K(N\tau)$ is approximated by

\begin{equation}
    K(N\tau)
    =
    (N\tau)^2
    \operatorname{sinc}^2
    \left[
    \frac{N\tau}{2}
    \left(
    \omega_L-\frac{\pi}{\tau}
    \right)
    \right]
    \label{eq:filter_function}
\end{equation}

At resonance, the Larmor frequency and DD pulse interval satisfy $\omega_L = \frac{\pi}{\tau}$, and $K(N\tau)$ reduces to

\begin{equation}
    K(N\tau) = (N\tau)^2 = t_{\mathrm{sens}}^2
    \label{eq:filter_on_resonance}
\end{equation}


Let $S_{\mathrm{bg}}$ denote the differential photon-count signal in the DD background (baseline). At the NMR resonance, the signal $S_{\mathrm{DD}}$ is reduced to

\begin{equation}
    S_{\text{DD}}
    =
    S_{\mathrm{bg}} C_{\mathrm{DD}}(\tau)
    =
    S_{\mathrm{bg}}
    \exp\left[
    -AB_{\mathrm{rms}}^2
    \right]
    \label{eq:photon_signal}
\end{equation}

Consequently, under the photon-shot noise limit, the sensitivity can be obtained as





\begin{equation}
    \eta_{B_{\mathrm{rms}}}
    =
    \frac{\sigma_{\text{signal}}}{\left| \frac{\partial S_{\mathrm{DD}}}{\partial B_{\mathrm{rms}}} \right|} \sqrt{T_{\text{total}}}
    =
    \frac{
    \sqrt{F_0+F_1}
    }{
    2AB_{\mathrm{rms}}
    \left|S_{\mathrm{bg}}\right|
    \exp\left[
    -AB_{\mathrm{rms}}^2
    \right]
    }
    \sqrt{T_{\mathrm{total}}}.
    \label{eq:final_sensitivity}
\end{equation}

In the experiment, $T_{\text{total}}$ is the total acquisition time for the two-sided measurement and is therefore written as $2 \times T_{\text{sequence}}$, where $T_{\text{sequence}}$ is the duration of one DD sequence including the overhead. The derivation assumes that the DD background $S_{\mathrm{bg}}$ is accurately determined from a fitted background envelope or sufficient off-resonance averaging. A full discussion of sensitivity is available in Ref.~\cite{phamNMRTechniqueDetermining2016}.

\subsection*{Effective depth of the NV ensemble}

The target sample is assumed to have a uniform spin density $\rho$ and to occupy a semi-infinite half-space above the diamond surface. For an NV center at a depth $d_{\mathrm{NV}}$ below the sample surface, $B_{\mathrm{rms}}$ can be expressed as \cite{abendrothSingleNitrogenVacancyNMRAmineFunctionalized2022}

\begin{align}
B_{\text{rms}}^{2}
&=
\int_{d_{\text{NV}}}^{\infty}
\int_{-\infty}^{\infty}
\int_{-\infty}^{\infty}
\rho
\left(
\frac{\mu_{0}}{4\pi}
\frac{\hbar \gamma_I}{2}
\right)^2
\frac{9}{\left(x^2+y^2+z^2\right)^5}
\nonumber \\
&\quad \times
\left[
\left((x\sin\theta+z\cos\theta)y\right)^2
\right.
\nonumber \\
&\qquad \left.
+
\left((x\sin\theta+z\cos\theta)(x\cos\theta-z\sin\theta)\right)^2
\right]
\, dx\,dy\,dz,
\label{eqS4}
\end{align}
where $x$, $y$, and $z$ are the spatial coordinates of the target spin, $\mu_0$ is the vacuum permeability, $\hbar$ is the reduced Planck constant, and $\gamma_I$ is the gyromagnetic ratio of the target spins.
For the NV orientation along [111], $\theta$ is zero, and the integral can be evaluated analytically as

\begin{equation} \label{eqS5}
    B_{\mathrm{rms}}^{2} = \frac{\gamma_I^2 \hbar^2 \mu_0^2 \rho}{256\pi d_{\text{NV}}^3}
\end{equation}









\subsection*{Quantum sensing using R0 and R2 in sample S1}

\begin{figure}[bt]
	\centering
    \includegraphics{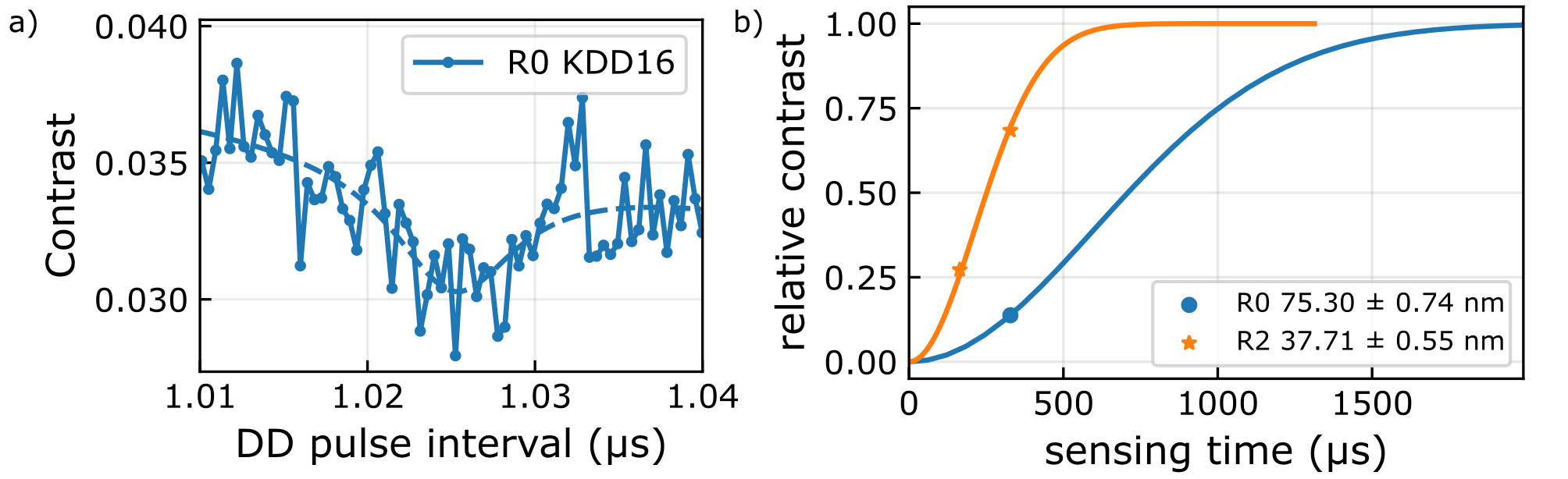}
	\caption{\textbf{Magnetic sensitivity and depth estimations.}
    \textbf{a)} The KDD16 spectrum measured with R0 of sample S1.
    \textbf{b)} The depth estimations of R0 and R2 in sample S1.
    }
	\label{figS3}
\end{figure}

All data are shown in Fig.~\ref{fig5} of the main text and Fig.~S\ref{figS3}. We perform DD measurements in regions R0 and R2 of sample S1 to demonstrate the suitability of these NV layers for quantum sensing. Using all experimental parameters, the sensitivity is estimated to be \SI{47.39}{nT/\sqrt{\text{Hz}}} for R0 and \SI{17.66}{nT/\sqrt{\text{Hz}}} for R2. The calculated effective depths of the NV ensembles are \SI{75.30}{nm} for R0 and \SI{37.71}{nm} for R2, using a proton spin density of $\rho = \SI{60}{spins / nm^3}$ \cite{loretzNanoscaleNuclearMagnetic2014a}.
For an NV ensemble, $d_{\mathrm{NV}}$ is the effective depth representing the entire ensemble. Notably, the effective depths for both regions do not lie at the centers of the corresponding thicknesses (\SI{217}{nm} and \SI{53}{nm} for R0 and R2, respectively). This can be possibly explained by the dependence of the effective depth on the measurement time, because the coherence properties of individual NV centers depend strongly on their depth. Thus, for a longer measurement time, the effective depth is weighted more strongly toward deeper NV centers. In principle, different values of $d_{\mathrm{NV}}$ and $\eta_{\mathrm{B}}$ can be achieved by tuning the measurement time. In our experiments, the parameters are chosen to make the measurements straightforward to perform and mutually comparable. Better sensitivity may be obtained by optimizing the measurement parameters. Nevertheless, the results show that the NV layers in both R0 and R2 are suitable for quantum-sensing applications and potentially for studies of many-body physics in which the interacting spins are tuned.

\newpage
\bibliographystyle{naturemag}
\bibliography{SI_references}